# A revisit of generalized scaling of forced turbulence through flux analysis

Wei Zhao

*State Key Laboratory of Photon-Technology in Western China Energy, International Scientific and Technological Cooperation Base of Photoelectric Technology and Functional Materials and Application, Institute of Photonics and Photon-technology, Northwest University, Xi'an 710127, China*

**Abstract**

In this investigation, we theoretically studied the transports of kinetic energy and scalar variance in turbulence driven by a scalar-based volume force in $\boldsymbol{M}\nabla^{\beta}s'$ form associated with scalar fluctuations $s'$ in wavenumber space relies on flux conservation equation. The equation has one real solution and two complex solutions, which lead to four different cascade processes, including inertial subrange (constant fluxes of kinetic energy and scalar variance), CEF subrange (quasi-constant flux of kinetic energy), CSF subrange (quasi-constant flux of scalar variance), and a new subrange with both non-constant fluxes of kinetic energy and scalar variance in addition to dissipation subrange.

$\beta$ controls the cascade processes and the scaling exponents. For the real solution, in the CEF subrange, $\xi_u$ is always -5/3, while $\xi_s = -(6\beta + 1)/3$. In the CSF subrange, $\xi_u = (4\beta - 11)/5$ and $\xi_s = -(2\beta + 7)/5$ which are both consistent with the theory of Zhao and Wang (2021). Relying on $\beta$, the transport of kinetic energy and scalar variance can be distinguished as four cases. (1) When $\beta < 3/2$ (except $\beta = 2/3$), the CEF and CSF subranges are coexisted, with the former located on the lower wavenumber side of the latter. At $\beta = 2/3$, a new inertial subrange with both $\xi_u$ and $\xi_s$ equal to -5/3 is present. (2) When $3/2 \leq \beta < 2$, only the CEF subrange is predicted. (3) At $\beta = 2$, special and singular exponents of $\xi_u = -1$, $\xi_s = -3$, $\lambda_u = 1$, and $\lambda_s = -1$ can be found, if $MN = 1$. Otherwise, a CSF subrange is predicted. (4) When $2 < \beta \leq 4$, only the CSF subrange is predicted.

For the 2nd solution (one complex solution), when $\beta < 3/2$, there are no CEF and CSF subranges. When $\beta \geq 3/2$, a single CEF subrange is predicted. While for the 3rd solution (the other complex solution), there is always a CSF subrange at $\beta < 3/2$. However, when $\beta \geq 3/2$, the flux of kinetic energy and scalar variance become non-constant.

Thus, a complete transport picture of both kinetic energy and scalar variance has been established for the type of forced turbulence, which unifies stratified turbulence, turbulent thermal convection, electrokinetic turbulence, etc.

## 1. Introduction

How the energy and scalar are transported along with scales in scalar-based forced turbulence are still unsolved problems and might remain far from understood for a long while. Even in stratified turbulence that has been investigated for over half a century, this problem still remains debated. In the 1950s, Bolgiano (1959) and Obukhov (1959) separately advanced the celebrated BO59 law which can be derived by dimensional analysis on the basis of a constant scalar flux of temperate inferred from Benzi et al (1996). The cascade of kinetic energy and temperature variance can both be separated into a series of subranges from large to small scales in sequence, including large scale injection subrange, buoyancy-driven subrange (in stratified turbulence or turbulent thermal convection (Niemela et al. 2000, Lohse & Xia 2010, Kumar et al. 2014)), inertial subrange and dissipation subrange etc. In the buoyancy-driven subrange of stably stratified turbulence, the power spectra of velocity and temperature have the following scaling expressions (Davidson 2013, Alam et al. 2019)

$$E_u(k) \sim \varepsilon_T^{2/5} f_{VB}^{4/5} k^{-11/5} \tag{1.1a}$$

$$E_T(k) \sim \varepsilon_T^{4/5} f_{VB}^{-2/5} k^{-7/5} \tag{1.1b}$$

where $f_{VB} = \sqrt{\frac{g}{\langle \rho \rangle} \left|\frac{d\rho_0}{dz}\right|}$ is Väisälä–Brunt frequency, with $\rho_0$, $\langle \rho \rangle$ and $g$ are background density, mean density with perturbations and gravity respectively. While in the inertial subrange, they got

$$E_u(k) \sim \varepsilon_u^{2/3} k^{-5/3} \tag{1.2a}$$

$$E_T(k) \sim \varepsilon_u^{-1/3} \varepsilon_T k^{-5/3} \tag{1.2b}$$

One essential problem is what the conservative law (or constant quantity) is in the transport of kinetic energy and scalar variance in the scalar-based forced turbulence.

In BO59 law, in the buoyancy-driven subrange, the flux of temperature variance is hypothesized to be constant, while in the inertial subrange, both the fluxes of kinetic energy and temperature variance are constant. The constant flux of scalar variance is also adopted by Zhao and Wang to establish the cascade frameworks of electrokinetic turbulence (Zhao & Wang 2017, Zhao & Wang 2019) and a general form of forced turbulence with $\nabla^n s$-type volume forces (Zhao & Wang 2021), where $s$ is a control scalar for active turbulent transport process. A series of new scaling phenomena, e.g. -7/5 velocity spectrum and -9/5 scalar spectrum have been predicted in electrokinetic turbulence. Nevertheless, the hypothesis of the constant flux of scalar variance results in an unlimited perturbation of flow on small scales when $n > 4$ (Zhao & Wang 2021). The statistical equilibrium system of scalar turbulence is accordingly believed to be collapsed.

However, if we take a look at the investigations above, it is interesting to see three different cascade processes, depending on the fluxes of kinetic energy and scalar variance. In the inertial subrange, both the fluxes of kinetic energy and scalar variance are constant. In the VFD subrange, a constant flux of scalar variance with a non-constant flux of kinetic energy has been reported. While in the dissipation subrange, both the fluxes of kinetic energy and scalar variance are non-constant. It is curious to ask if there should be a subrange with constant kinetic energy and non-constant scalar variance.

In 2019, Alam et al. (2019) revisited BO59 law by combing the flux of kinetic energy and potential energy (which is essentially related to temperature variance). They suggest considering the summation of kinetic energy flux ($\Pi_u(k)$) and temperature variance flux ($\Pi_T(k)$) to be constant, instead of considering each individual to be constant, in the subranges $k \ll k_\eta$ ($k_\eta$ is Kolmogorov wavenumber). A -5/3 slope of $E_u(k)$ was found at the lower wavenumber side of the buoyancy-driven subrange, where the spectra of temperature variance, i.e. $E_T(k)$, has a slope of -1/3. In this subrange, only $\Pi_u(k)$ is approximately constant. Beyond this subrange, the scaling subrange predicted in BO59 law is found numerically, where only $\Pi_T(k)$ is approximately constant. The existence of the former subrange (-1/3 spectra of $E_T(k)$) is not supported by the previous investigations in stratified turbulence. However, the investigations of Alam et al (2019) are highly inspiring, as they successfully show the possible existence of constant $\Pi_u(k)$ subrange and may complete the jigsaw of cascade in scalar-based forced turbulence.

In this investigation, we extend the analysis of Zhao and Wang (2021) and revisit the model of generally forced turbulence with $\nabla^\beta s'$-type volume forces, to established a universal theory for general conservative law in the turbulence with stratification-like scalar background. Besides the influence of derivation order $\beta$, we also take the selection of conservative equation solutions into account. Rich information of characteristic microscales has been provided accordingly.

## 2. Theory

In this investigation, we revisit a general model of scalar-based forced turbulence (Zhao & Wang 2021) with stratified scalar features, by analogy with the analytical method of Alam et al (2019). The turbulent model can be described with the following equations

$$\frac{D\boldsymbol{u}}{Dt} = -\frac{1}{\rho}\nabla p + \nu\nabla^2\boldsymbol{u} + \boldsymbol{M}\nabla^\beta s' \tag{2.1a}$$

$$\frac{Ds'}{Dt} = -\boldsymbol{N}\cdot\boldsymbol{u} + D_s\nabla^2 s' \tag{2.1b}$$

$$\nabla\cdot\boldsymbol{u} = \boldsymbol{0} \tag{2.1c}$$

where $\rho$ is the fluid density, $\boldsymbol{u}$ denotes velocity vector, $p$ is pressure, $s'$ is any scalar fluctuations that control the volume force. $\boldsymbol{M}\nabla^\beta s'$ is a volume force term which is determined by the scalar field $s'$, with $\boldsymbol{M}$ being a certain dimensional vector associated with the physical field and $\nabla^\beta s'$ being a scalar function associated with a specific $\beta^{\text{th}}$-order derivative of the scalar $s'$. $\boldsymbol{N}$ is a vector related to scalar $s$ to characterize the feature of stratified scalar background, and $\nu$ and $D_s$ are the kinematic viscosity and diffusivity of the scalar respectively. Eqs. (2.1a-b) require $\boldsymbol{M}$ and $\boldsymbol{N}$ have the dimensions of $U^2 L^{\beta-1} S^{-1}$ and $S/L$, where $U$, $L$ and $S$ represent the dimensions of velocity, length and scalar respectively. For buoyancy-driven turbulence, let $s' = \rho' g f_{VB}^{-1}\langle\rho\rangle^{-1}$ being density fluctuations, $\beta = 0$, $\boldsymbol{M} = \boldsymbol{N} = -f_{VB}\hat{\boldsymbol{z}}$, the control equations (2.1a-b) are consistent with Alam et al. (2019). For electrokinetic turbulence with a quasi-stratified mean scalar field, we can let $\beta = 1$, $\boldsymbol{M} = -\varepsilon E^2\hat{\boldsymbol{y}}/\rho\langle\sigma\rangle$, $\boldsymbol{N} = \nabla\langle\sigma\rangle$, where $\sigma$ is electric conductivity, $E$ is electric field in $y$-direction ($\hat{\boldsymbol{y}}$) that perpendicular to initial interface of electric conductivity (Wang et al. 2014, Wang et al. 2016). For one-dimensional spectra, according to Verma (2004, Verma 2018)

$$E_u(k)dk = \sum_{k<|\boldsymbol{k}'|\leq k+dk}\frac{1}{2}|u(\boldsymbol{k}')|^2 \tag{2.2a}$$

$$E_s(k)dk = \sum_{k<|\boldsymbol{k}'|\leq k+dk}\frac{1}{2}|s'(\boldsymbol{k}')|^2 \tag{2.2b}$$

$$\Pi_u(k_0) = \sum_{|\boldsymbol{k}|>k_0}\sum_{|\boldsymbol{m}|\leq k_0}\text{Im}\{[\boldsymbol{k}\cdot\boldsymbol{u}(\boldsymbol{n})][\boldsymbol{u}(\boldsymbol{m})\cdot\boldsymbol{u}^*(\boldsymbol{k})]\} \tag{2.2c}$$

$$\Pi_s(k_0) = \sum_{|\boldsymbol{k}|>k_0}\sum_{|\boldsymbol{m}|\leq k_0}\text{Im}\{[\boldsymbol{k}\cdot\boldsymbol{u}(\boldsymbol{n})][s'(\boldsymbol{m})s'^*(\boldsymbol{k})]\} \tag{2.2d}$$

where $E_u(k)$ is kinetic energy spectrum, $E_s(k)$ is scalar energy spectrum, $\Pi_u(k_0)$ and $\Pi_s(k_0)$ are the fluxes of kinetic energy and scalar variance in spectral space respectively. $\boldsymbol{k} = \boldsymbol{m} + \boldsymbol{n}$. Let $E_u(\boldsymbol{k}) = \frac{1}{2}|\boldsymbol{u}(\boldsymbol{k})|^2$ and $E_s(\boldsymbol{k}) = \frac{1}{2}|s'(\boldsymbol{k})|^2$, if we consider the contribution of scalar-based volume force (SVF) to scalar transport, under a statistical equilibrium state, we have (Davidson 2013, Alam et al. 2019)

$$T_u(\boldsymbol{k}) + F_s(\boldsymbol{k}) - D_u(\boldsymbol{k}) = 0 \tag{2.3a}$$

$$T_s(\boldsymbol{k}) - F_A(\boldsymbol{k}) - D_s(\boldsymbol{k}) = 0 \tag{2.3b}$$

$T_u(\mathbf{k})$ and $D_u(\mathbf{k})$ are the nonlinear kinetic energy transfer rate and dissipation rate, respectively. $T_s(\mathbf{k})$ and $D_s(\mathbf{k})$ are the nonlinear transfer rate of scalar variance and scalar dissipation rate, respectively. $F_s(\mathbf{k})$ denotes the energy feeding rate by SVF and $F_A(\mathbf{k})$ is the scalar feeding rate by bulk components. These quantities can be expressed as

$$T_u(\mathbf{k}) = \sum_m \text{Im}\{[\mathbf{k}\cdot\mathbf{u}(\mathbf{n})][\mathbf{u}(\mathbf{m})\cdot\mathbf{u}^*(\mathbf{k})]\} \tag{2.4a}$$

$$T_s(\mathbf{k}) = \sum_m \text{Im}\{[\mathbf{k}\cdot\mathbf{u}(\mathbf{n})][s'(\mathbf{m})s'^*(\mathbf{k})]\} \tag{2.4b}$$

$$F_s(\mathbf{k}) = \text{Re}\big[(\mathbf{M}\nabla^\beta s')(\mathbf{k})\cdot\mathbf{u}^*(\mathbf{k})\big] = \text{Re}\big[s'(\mathbf{k})k^\beta \mathbf{M}\cdot\mathbf{u}^*(\mathbf{k})\big] \tag{2.4c}$$

$$F_A(\mathbf{k}) = \text{Re}[s'(\mathbf{k})\mathbf{N}\cdot\mathbf{u}^*(\mathbf{k})] \tag{2.4d}$$

$$D_u(\mathbf{k}) = 2\nu k^2 E_u(\mathbf{k}) \tag{2.4e}$$

$$D_s(\mathbf{k}) = 2D_s k^2 E_s(\mathbf{k}) \tag{2.4f}$$

Eq. (2.4c) is not a strict mathematically derivation, but a dimensional one. Its explicit form should be derived based on the definition of the operator $\nabla^\beta$. Even though, the major purpose of applying the $\mathbf{M}\nabla^\beta s'$ volume force in this investigation can be seen in Eq. (2.4c). Generally, we can assume $F_s(\mathbf{k})$ have a form of $\text{Re}[s'(\mathbf{k})W(k)\cdot\mathbf{u}^*(\mathbf{k})]$, which is established on the basis of the cospectra $s'(\mathbf{k})\mathbf{u}^*(\mathbf{k})$ and a coupling function $W(k)$ which can be expanded polynomially as $W(k) = \sum_{n=0}^{\infty} a_n k^n$ (where $n$ is an integer). Then, the contribution of the dominant $n^{\text{th}}$ order term can be estimated through $\text{Re}[s'(\mathbf{k})a_n k^n \cdot \mathbf{u}^*(\mathbf{k})]$. However, if there exist some neigthboring orders (e.g. $a_0 + a_1 k$) that all exhibit important contributions to the energy feeding rate, we may use a simple power-law equation $a_\beta k^\beta$ with a non-integer exponent to approximate the polynomial terms. Therefore, by taking into account both integer and non-integer orders above, and distinguishing the influence of the dominant order on the transport of kinetic energy and scalar variance, the $\mathbf{M}\nabla^\beta s'$ volume force is adopted in this investigation.

Combining Eqs. (2.2c-d) with (2.4a-b), it is obtained,

$$\Pi_u(k_0) = -\sum_{|\mathbf{k}|\leq k_0} T_u(\mathbf{k}),\ \Pi_s(k_0) = \sum_{|\mathbf{k}|\leq k_0} T_s(\mathbf{k}) \tag{2.5}$$

After simple processing on Eqs. (2.2) to (2.4), we have

$$0 = -\frac{\text{d}}{\text{d}k}\Pi_u(k) + F_s(k) - D_u(k) \tag{2.6a}$$

$$0 = -\frac{\text{d}}{\text{d}k}\Pi_s(k) - F_A(k) - D_s(k) \tag{2.6a}$$

where

$$F_s(k)\text{d}k = \sum_{k<|\mathbf{k}'|\leq k+\text{d}k} \text{Re}\big[s'(\mathbf{k}')k'^\beta \mathbf{M}\cdot\mathbf{u}^*(\mathbf{k}')\big] \tag{2.7a}$$

$$F_A(k)\text{d}k = \sum_{k<|\mathbf{k}'|\leq k+\text{d}k} \text{Re}[s'(\mathbf{k}')\mathbf{N}\cdot\mathbf{u}^*(\mathbf{k}')] \tag{2.7b}$$

$$D_u(k)\text{d}k = 2\nu \sum_{k<|\mathbf{k}'|\leq k+\text{d}k} k'^2 E_u(\mathbf{k}') \tag{2.7c}$$

$$D_s(k)\text{d}k = 2D_s \sum_{k<|\mathbf{k}'|\leq k+\text{d}k} k'^2 E_s(\mathbf{k}') \tag{2.7d}$$

When $\text{d}k \to 0$, it is approximately having

$$F_s(k) = \text{Re}\big[s'(\mathbf{k})k^\beta \mathbf{M}\cdot\mathbf{u}^*(\mathbf{k})\big] \tag{2.8a}$$

$$F_A(k) = \text{Re}[s'(\mathbf{k})\mathbf{N}\cdot\mathbf{u}^*(\mathbf{k})] \tag{2.8b}$$

$$D_u(k) = 2\nu k^2 E_u(k) \tag{2.8c}$$

$$D_s(k) = 2D_s k^2 E_s(k) \tag{2.8d}$$

In the inertial subrange, where the influence of forcing and dissipation is negligible, we have

$$\frac{d}{dk} \Pi_u(k) = 0 \tag{2.9a}$$

$$\frac{d}{dk} \Pi_s(k) = 0 \tag{2.9b}$$

Or

$$\Pi_u(k) = \text{const along } k \tag{2.10a}$$

$$\Pi_s(k) = \text{const along } k \tag{2.10b}$$

The inertial subrange must have constant $\Pi_u$ and $\Pi_s$. This is a special solution of Eq. (2.12) as can be seen later.

In the VFD subrange, the dissipation terms of kinetic energy and scalar variance are ignored, whereas

$$\frac{d}{dk} \Pi_u(k) = F_s(k) \tag{2.11a}$$

$$\frac{d}{dk} \Pi_s(k) = -F_A(k) \tag{2.11b}$$

Thus, if $\boldsymbol{M}$ is parallel to $\boldsymbol{N}$, from Eqs. (2.8a-b) and (2.11a-b), easily we get

$$F_s(k) - \frac{M}{N} F_A(k) k^\beta = 0 = \frac{d}{dk} \Pi_u(k) + \frac{M}{N} k^\beta \frac{d}{dk} \Pi_s(k) \tag{2.12}$$

where $M = |\boldsymbol{M}|$ and $N = |\boldsymbol{N}|$ respectively. According to Alam et al (2019),

$$E_u(k) = u_k^2 / k \tag{2.13a}$$

$$E_s(k) = s_k^2 / k \tag{2.13b}$$

$$\Pi_u(k) = k u_k^3 \tag{2.13c}$$

$$\Pi_s(k) = k s_k^2 u_k \tag{2.13d}$$

Substituting $\Pi_u(k)$ and $\Pi_s(k)$ into the Eq. (2.12) above,

$$\frac{d}{dk} k u_k^3 + \frac{M}{N} k^\beta \frac{d}{dk} k s_k^2 u_k = 0 \tag{2.14}$$

When $\beta = 0$ and $M = N = -f_{VB}$, Eq. (2.14) is equivalent to Eq. (2.15) in the manuscript of Alam et al. (2019) for stratified turbulence. Considering in VFD subrange, the flow is driven by the $\boldsymbol{M}\nabla^\beta s'$ type volume force, dimensionally $k u_k^2 = k^\beta M s_k$, then Eq. (2.14) becomes

$$u_k + \frac{(3-2\beta)}{MN} k^{2-\beta} u_k^3 + \left(3k + \frac{5}{MN} k^{3-\beta} u_k^2\right) \frac{du_k}{dk} = 0 \tag{2.15}$$

Eq. (2.15) is a universal equation for the VFD subrange in the forced turbulence driven by $\boldsymbol{M}\nabla^\beta s'$ type volume force. There are three solutions in Eq. (2.15), including one real solution and two complex solutions. Here, all the three solutions are taken into account, say $u_{k,i}^2 = u_{k,i} \cdot u_{k,i}^*$ with $u_{k,i}^*$ being the complex conjugate of $u_{k,i}$ which is the ith solution of Eq. (2.15). Thus, we can simply rewrite Eq. (2.13a-d) as

$$E_u(k) = k^{-1}(u_k \cdot u_k^*) \sim k^{\xi_u} \tag{2.16a}$$

$$E_s(k) = M^{-2}k^{3-2\beta}E_u^2(k) \sim k^{\xi_s} \tag{2.16b}$$

$$\Pi_u(k) = k^{5/2}E_u^{3/2}(k) \sim k^{\lambda_u} \tag{2.16c}$$

$$\Pi_s(k) = M^{-2}k^{\frac{11}{2}-2\beta}E_u^{5/2}(k) \sim k^{\lambda_s} \tag{2.16d}$$

where $\xi_u, \xi_s, \lambda_u, \lambda_s$ denote the scaling exponents of $E_u, E_s, \Pi_u$ and $\Pi_s$ respectively.

## 3. Numerical results

Eq. (2.15) can be solved numerically after settling $\beta$, $M$ and $N$. Nevertheless, to directly simulate Eq. (2.15) in a wide wavenumber range, e.g. from $10^{-5}$ to $10^{25}$, the cost of computation is unaffordable in linear wavenumber space. Therefore, we made a transform from the linear wavenumber space to the log space (or decades), i.e. $k = 10^q$, thus Eq. (2.15) becomes

$$u_k + \frac{(3-2\beta)}{MN}10^{q(2-\beta)}u_k^3 + \frac{1}{\ln 10}\left[3 + \frac{5}{MN}10^{q(2-\beta)}u_k^2\right]\frac{du_k}{dq} = 0 \tag{3.1}$$

In the $q$ space, this equation can be solved simply with

$$u_{k,i,m} + \frac{(3-2\beta)}{MN}10^{q_m(2-\beta)}u_{k,i,m}^3 + \frac{1}{\ln 10}\left[3 + \frac{5}{MN}10^{q_m(2-\beta)}u_{k,i,m}^2\right]\frac{u_{k,i,m}-u_{k,i,m-1}}{q_m-q_{m-1}} = 0 \tag{3.2}$$

for $m = 2, 3, \ldots, m_{max}$, where $m_{max}$ is the maximum index of the decade $q_m$ in this investigation. $u_{k,i,m}(q_m)$ denotes the $u_{k,i}$ at $q_m$, or the equivalent wavenumber $k_m = 10^{q_m}$. Apparently, the solution of Eq. (3.1) relies on an initial $u_{k,i}$ at $m=1$, i.e. $u_{k,i,1}$. Considering the total kinetic energy $\int E_u(k)dk$ is inevitably determined by $MN$, the distribution of $u_k$ (including $u_{k,i,1}$) relies on $MN$ as well. The influence of $\beta$, $MN$ and the initial value $u_{k,1,1}$ of the first (real) solution will be comprehensively discussed in the following sections.

### *3.1 Influence of $\beta$*

According to Eq. (2.15), it is explicitly seen that $\beta$ is a crucial parameter in determining the transport of kinetic energy and scalar variance. Zhao and Wang (2021) theoretically predicted that $\beta$ should be no more than 4, otherwise, the statistical equilibrium of the system can be broken, as a result of unlimited accumulation of kinetic energy at small scales. Although the conclusion is attributed to a constant flux of scalar variance which is absent in some cases, the restriction (i.e. $\beta \leq 4$) is still applicable as discussed in section 3.2.

**(1) $\beta < 3/2$**

Generally, when $\beta < 3/2$ (except $\beta = 2/3$), the VFD subrange can always be separated into two additional subranges (Figure 1). One has a quasi-constant flux of kinetic energy (say "CEF" subrange), and the other has a quasi-constant flux of scalar variance (say "CSF" subrange). The two subranges intersect with each other at a critical wavenumber $k_{ci}$.

[(i) *CEF subrange*](#)

Figure 1(a) shows the variation of $E_u(k)$ versus $\beta$ in a large wavenumber range. When $\beta < 3/2$ (except $\beta = 2/3$), an inertial subrange with $\xi_u = -5/3$ and a quasi-constant flux of kinetic energy with the slope $\lambda_u \sim 0$ (Figure 1(c)) can be found in the lower wavenumber range of $k \ll k_{ci}$. Since $\lambda_u$ is not exactly zero, but has a very small (~1/30 or smaller) magnitude which is negligible, we call it "quasi-constant". This explains why $\Pi_s$ can be non-constant (Figure 1(d)) according to Eq. (2.12).

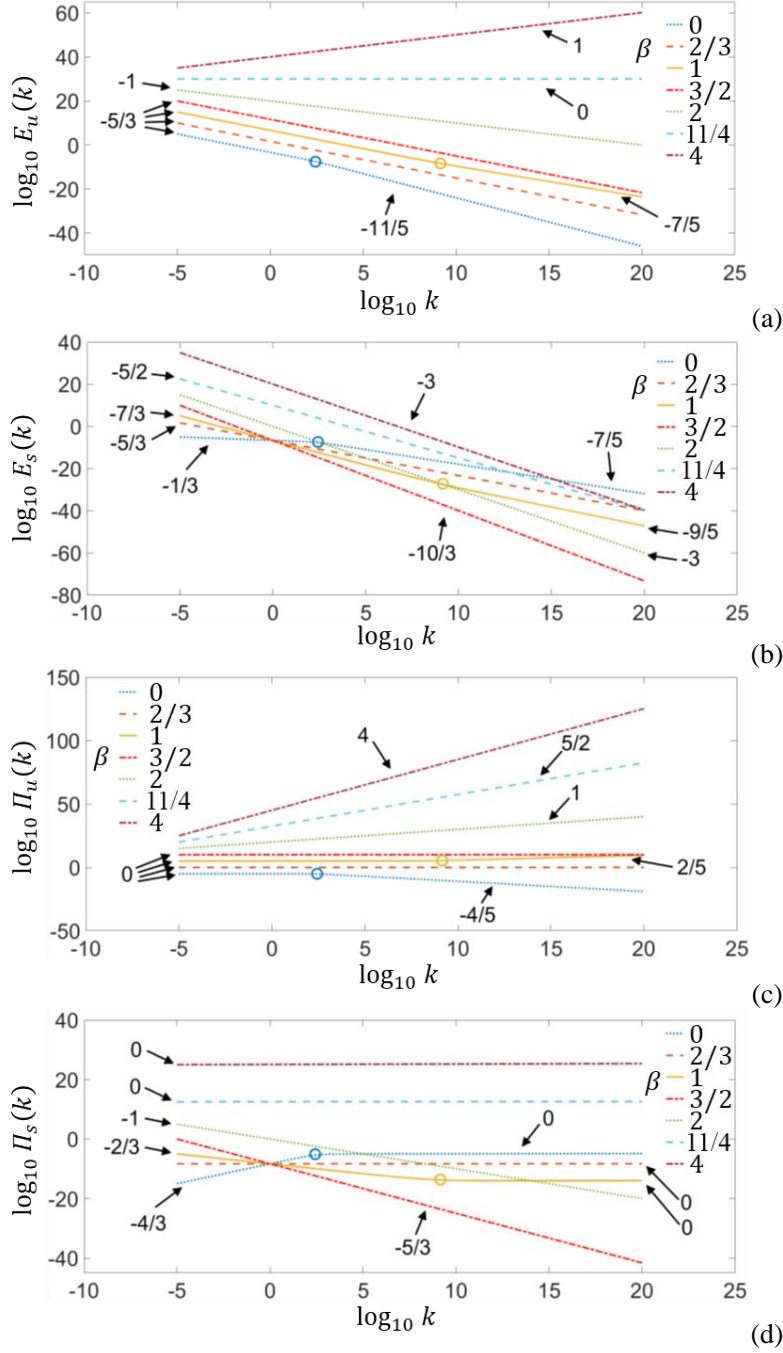

Figure 1. Influence of $\beta$ on the transport of kinetic energy and scalar variance, where $MN = 1$ and only the first solution (real solution), i.e. $u_{k,1}$, has been taken into account. The initial value $u_{k,1,1} = 1$. (a) Power spectra of kinetic energy $E_u$. Each curve has been shifted by 5 vertically for better clarity. (b) Power spectra of scalar variance $E_s$. (c) Flux of kinetic energy $\Pi_u$. Each curve has been shifted by 5 vertically for better clarity. (d) Flux of scalar variance $\Pi_s$.

In the CEF subrange, $E_s(k)$ shows a new scaling behaviour which relies on $\beta$, as plotted in Figure 1(b). $\xi_s$ is linearly related to $\beta$ as

$$\xi_s = -(6\beta + 1)/3 \qquad (3.3)$$

To the best of the author's knowledge, this scaling law of scalar spectra has not been predicted by other theoretical analyses or numerical simulations. In the CEF subrange, the flux of scalar variance has a scaling exponent as

$$\lambda_s = (4 - 6\beta)/3 \qquad (3.4)$$

Since $\lambda_u = 0$ and $\lambda_s = (4 - 6\beta)/3$, according to Eqs. (2.11), we thus have $F_s(k) \sim k^{-1}$ and $F_A(k) \sim k^{(1-6\beta)/3}$.

(ii) *CSF subrange*

When $\beta < 3/2$ (except $\beta = 2/3$), as shown in Figure 1(d), the CSF subrange where $\lambda_s \sim 0$ is found located at the higher wavenumber range of the CEF subrange, i.e. $k \gg k_{ci}$. From the numerical calculation, it is found $\xi_u$ in the CSF subrange follows

$$\xi_u = (4\beta - 11)/5 \qquad (3.5)$$

This is explicitly consistent with the theory of Zhao and Wang (2021), which assumes $\Pi_s(k)$ is constant. The corresponding scaling exponents of $\Pi_u(k)$ becomes

$$\lambda_u = (6\beta - 4)/5 \qquad (3.6)$$

Since $\lambda_s$ is approximately zero (below 1/40), $\Pi_s(k)$ is quasi-constant in the CSF subrange. Here, $\xi_s$ is linearly related to $\beta$ as

$$\xi_s = -(2\beta + 7)/5 \qquad (3.7)$$

which is also consistent with the predictions by Zhao and Wang (2021). In the CSF subrange, since $\lambda_u = (6\beta - 4)/5$ and $\lambda_s = 0$, according to Eqs. (2.11), we have $F_s(k) \sim k^{(6\beta-9)/5}$ and $F_A(k) \sim k^{-1}$.

(iii) *Special case at $\beta = 2/3$*

From Figure 1, it is interesting to see a special case at $\beta = 2/3$, under which $\xi_u = \xi_s = -5/3$ and $\lambda_u = \lambda_s = 0$ in the VFD subrange. This means, at this $\beta$, there exists a delicate balance among the volume force, the transport of kinetic energy and scalar variance, which makes both kinetic energy and scalar variance experience inertial cascades from large to small scales. This result can be predicted if we rewrite Eq. (2.12) as $\frac{d}{dk}\Pi_u(k) + \frac{1}{MN}k^\beta \frac{d}{dk}\left[k^{\frac{4}{3}-2\beta}\Pi_u(k)^{\frac{5}{3}}\right] = 0$. To make the equation established, both $\Pi_u(k) = const$ and $k^{\frac{4}{3}-2\beta} = const$ are required. Accordingly, $\beta = 2/3$. The result indicates, on one hand, the direct observations of constant fluxes of kinetic energy and scalar variance, and the corresponding -5/3 slopes in the spectra of kinetic energy and scalar variance, are insufficient to ensure the existence of K41 law (Kolmogorov 1941, Frisch 1995) and Obukhov-Corrsin law (Obukhov 1949, Corrsin 1951) in scalar-based forced turbulence. On the other hand, it is possible to significantly extend the width of the -5/3 scaling subrange by generating forced turbulence with $\beta = 2/3$.

## (2) $3/2 \leq \beta < 2$

In the investigations of Zhao and Wang (2021), they found $\beta = 3/2$ is a special value on which the second-order structure function of velocity becomes irrelevant to scales, if CSF is hypothesized. However, their prediction is not accurate in the range of $3/2 \leq \beta <$

2. From Figure 1, it can be seen $\beta = 3/2$ is a critical value beyond which the coexistence of CEF and CSF subranges is broken. In the VFD subrange, only the CEF subrange is observed, where $\xi_u = -5/3$ and $\lambda_u = 0$. $\xi_s$ and $\lambda_s$ can be calculated from Eqs. (3.3) and (3.4).

**(3) $\beta = 2$**

$\beta = 2$ is a special and singular case in this investigation. In this case, no CEF or CSF subrange was observed. The scaling exponents are $\xi_u = -1, \xi_s = -3, \lambda_u = 1$, and $\lambda_s = -1$ respectively. All the derivatives of the scaling exponents, e.g. $d\xi_u/d\beta$ and $d\xi_s/d\beta$, are not available, since $\xi_u, \xi_s, \lambda_u$ and $\lambda_s$ are all discontinuous at $\beta = 2$. However, as illustrated in Section 4, $\beta = 2$ is a special case only when $MN = 1$. In any practical application, since $MN$ cannot be exactly unity, the special case may not be observed.

**(4) $2 < \beta \leq 4$**

When $\beta$ is further increased, the CSF subrange returns with $\lambda_s = 0$, as can be seen from Figure 1(d). In contrast, the CEF subrange disappears in the entire wavenumber range investigated. In this $\beta$ range, $\xi_u, \lambda_u$ and $\xi_s$ can be again calculated on the basis of Eqs. (3.5-3.7).

Generally, as shown in Figure 2, when $\beta < 3/2$ or $2 < \beta \leq 4$, the CSF subrange is observed in the wavenumber range, where the theoretical predictions of Zhao and Wang (2021) can be applied. When $3/2 \leq \beta < 2$ or $\beta = 2$ with $MN \neq 1$, the CEF subrange is predictable.

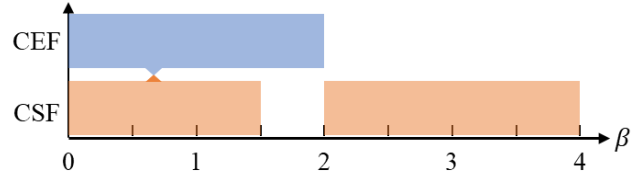

Figure 2 Relationship between $\beta$ and the energy or scalar transport subranges. The blue bar shows the CEF subrange and the orange one shows the CSF subrange. The small triangles highlight $\beta = 2/3$, where the CEF and CSF subranges overlap.

### 3.2 Influence of the magnitude of MN and $u_{k,1,1}$

$MN$ is a quantity to evaluate the forced convection effect in the scalar-based forced turbulence. According to Eq. (2.15), $M$ and $N$ have no difference in the scaling behaviour of $u_k$ and the related quantities. However, $M$ does affect the magnitudes of $E_s(k)$ and $\Pi_s$ from Eqs. (2.16b) and (2.16d). For better comparison, we change the magnitude of $MN$ through $M$, and keep $N = 1$ unchanged.

**(i)** When $\beta = 0$, from Figure 3(a), it can be seen in the CEF subrange, $\xi_u = -5/3$, $\xi_s = -1/3, \lambda_u = 0$ and $\lambda_s = 4/3$. While in the CSF subrange, $\xi_u = -11/5, \xi_s = -7/5$, $\lambda_u = -4/5$ and $\lambda_s = 0$. All these data are consistent with the numerical investigations of Alam et al (2019).

As $MN$ is increased, the critical wavenumber $k_{ci}$ increases rapidly (see Figure 3), which indicates the CEF subrange invades the CSF subrange and pushes it towards the higher wavenumber region. There is also no cross over observed between the VFD subrange and the inertial subrange in the investigated wavenumber region. Alam et al (2019) attributed the absence of inertial subrange to the insufficient strong disturbing from buoyancy. However, from Eq. (2.15), it can be predicted that if $\Pi_u$ and $\Pi_s$ are constants, we can only have a nontrivial solution of $u_k$ at $\beta = 2/3$. In other words, Eq. (2.15) that established for the VFD subrange is not applicable for the inertial subrange, and thus, it is reasonable that we cannot observe an intersection between the VFD and inertial subrange. A similar conclusion should also apply to the investigation of Alam et al (2019).

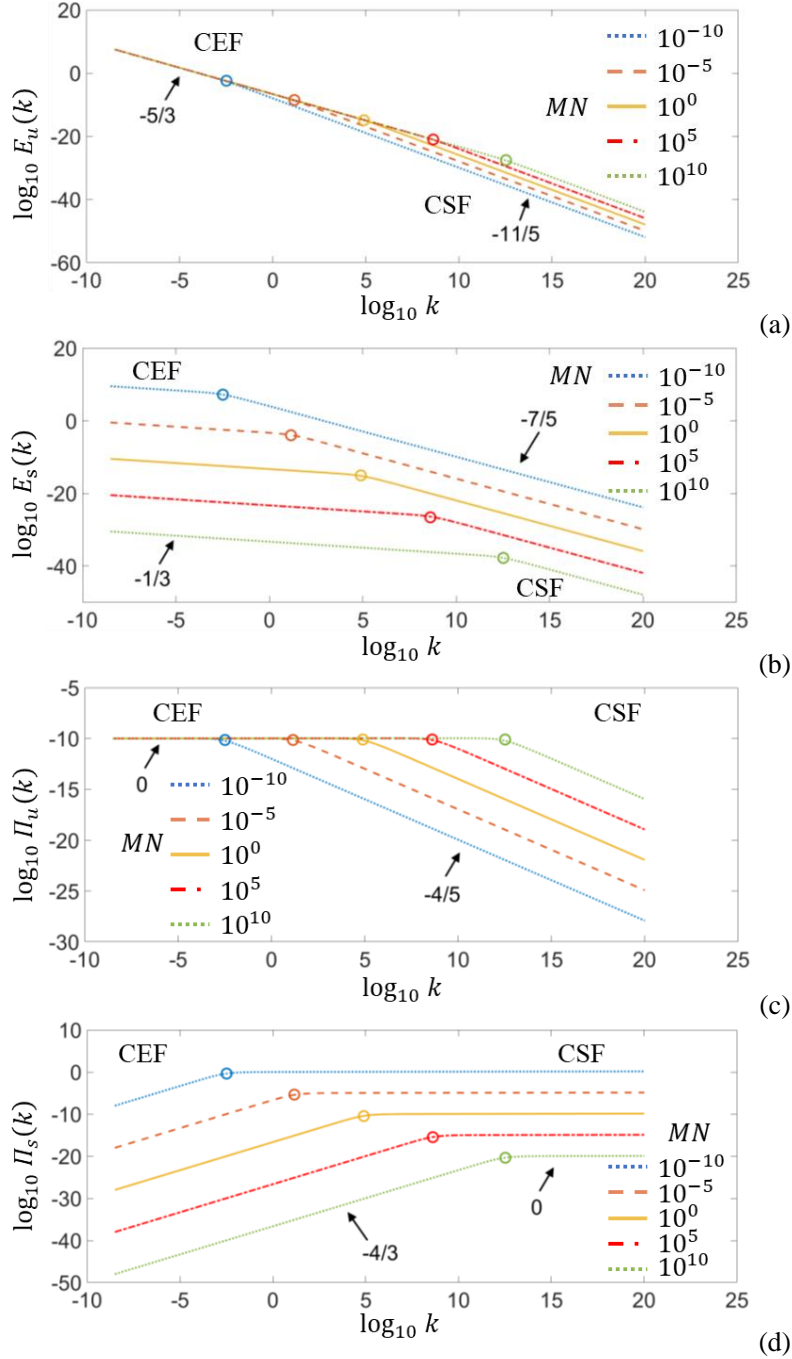

Figure 3 Influence of $MN$ on the transport of kinetic energy and scalar variance, where $\beta = 0$. Here, only the first solution (real solution), i.e. $u_{k,1}$, has been taken into account. The initial value $u_{k,1,1} = 1$. The colour circles denote the position of $k_{ci}$. (a) Power spectra of kinetic energy $E_u$. (b) Power spectra of scalar variance $E_s$. (c) Flux of kinetic energy $\Pi_u$. (d) Flux of scalar variance $\Pi_s$.

The increasing $MN$ also leads to a decrease of $E_s$ and $\Pi_s$. This result is apparently unreasonable and can be attributed to the fixed $u_{k,1,1}$ in the numerical calculation.

**(ii)** When $MN$ is changed, $u_{k,1,1}$ which is related to $E_u$, $E_s$, $\Pi_u$ and $\Pi_s$ curves at a large scale should also be changed accordingly. Or in other words, $u_{k,1,1}$ is a function of $MN$. In Figure 4, we show the influence of $u_{k,1,1}$ on the $E_u$, $E_s$, $\Pi_u$ and $\Pi_s$ curves at $\beta = 0$. As $u_{k,1,1}$ is increased, the critical wavenumber $k_{ci}$ decrease rapidly, which indicates the CSF subrange invades the CEF subrange and pushes it towards the lower wavenumber region. In the meanwhile, the magnitudes of $E_u$, $E_s$, $\Pi_u$ and $\Pi_s$ all increases with $u_{k,1,1}$.

**(iii)** The influence of $\beta$ on the relation between $MN$ and $k_{ci}$ has been plotted in Figure 5(a). In the log-log plot, all the curves exhibit a linear relationship, which indicates $k_{ci} \sim (MN)^\varphi$, where $\varphi = \varphi(\beta)$ is the scaling exponent relying on $\beta$. As $\beta$ is increased from 0 to 1, $\varphi$ increases rapidly from 0.75 (at $\beta = 0$) to 2.91 (at $\beta = 1$), as shown in Figure 5(c). The increase of $\varphi$ alongside $\beta$ is highly nonlinear and follows a quasi-exponential way. When $\beta = 5/4$, $\varphi$ is increased to an extraordinarily high value at 11.3. In other words, if $MN$ is doubled, $k_{ci}$ is increased 2400 times.

In Figure 5(b), we studied the influence of $u_{k,1,1}$ on $k_{ci}$. It is found that $k_{ci}$ changes with $u_{k,1,1}$ as $k_{ci} \sim u_{k,1,1}^\psi$, with $\psi = \psi(\beta)$ being another exponent relying on $\beta$. Relative to $\varphi$ which are always positive, $\psi$ are all negative with exactly the twice magnitudes of $\varphi$, as compared in Figure 5(c). As $\beta$ is increased from 0 to 1, $\varphi$ decreases rapidly from -1.5 (at $\beta = 0$) to -6 (at $\beta = 1$). As $\beta$ is further increased to $5/4$, $\psi$ is decreased to -22.7.

After taking the influence of both $MN$ and $u_{k,1,1}$ into account, we have $k_{ci} \sim (MN)^\varphi u_{k,1,1}^\psi$. The direct relationship between $u_{k,1,1}$ and $MN$ (or the related dimensionless parameters, e.g. Richardson number or Froude number in stratified turbulence (Maffioli & Davidson 2016, Howland et al. 2020, Okino & Hanazaki 2020)) relies on the physical problem and the dynamic system. It should be determined through either experiments or numerical simulations. For instance, if $u_{k,1,1} \sim (MN)^{1/3}$, considering $\psi \approx -2\varphi$ in this investigation, we simply have $k_{ci} = (MN)^{\varphi/3}$. In such a case, $k_{ci}$ increases with $MN$ and the CEF subrange moves toward higher wavenumber region, accompanied by a lifting of $E_u$. It means when the forced convection effect is stronger, larger kinetic energy and stronger velocity fluctuations are produced by forcing. Accordingly, it is more possible to induce a wider inertial subrange, through either direct or inverse cascade.

**(iv)** Zhao and Wang (2021) showed that when $\beta > 2/3$, the kinetic energy in the CSF subrange in the high wavenumber region is directly dissipated in the dissipation subrange. According to this investigation, the above conclusion is true in the range of $2/3 < \beta < 3/2$ and $2 < \beta \leq 4$. There exists a new cross over wavenumber $k_{us}$ which connects the CSF subrange and the dissipation subrange, where both the energy feeding rate and scalar feeding rate are balanced by the energy and scalar dissipation respectively. Accordingly, the following two conditions must be satisfied simultaneously

$$\begin{cases} F_s(k_{us}) = D_u(k_{us}) \\ F_A(k_{us}) = D_s(k_{us}) \end{cases} \tag{3.8}$$

After simple processing, Eq. (3.8) becomes

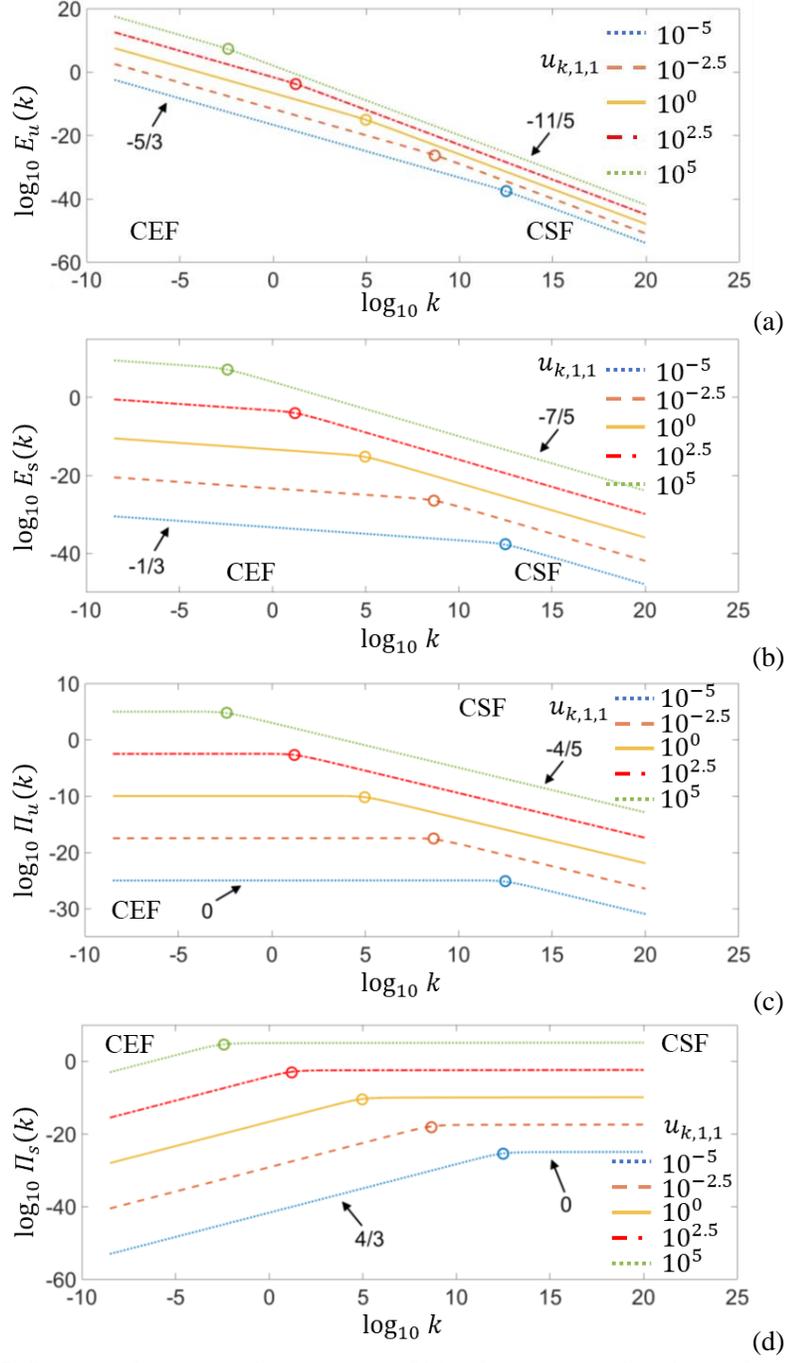

Figure 4 Influence of $u_{k,1,1}$ on the transport of kinetic energy and scalar variance, where $\beta = 0$ and $MN = 1$. The colour circles denote the position of $k_{ci}$. (a) Power spectra of kinetic energy $E_u$. (b) Power spectra of scalar variance $E_s$. (c) Flux of kinetic energy $\Pi_u$. (d) Flux of scalar variance $\Pi_s$.

$$\begin{cases} M[E_s^{1/2}(k_{us})E_u^{-1/2}(k_{us})k_{us}^{\beta-2}] = 2\nu \\ N[E_s^{-1/2}(k_{us})E_u^{1/2}(k_{us})k_{us}^{-2}] = 2D_s \end{cases} \quad (3.9)$$

which subsequently leads to

$$k_{us} = \left(\frac{MN}{4\nu D_s}\right)^{\frac{1}{4-\beta}} \quad (3.10)$$

For instance, in EK turbulence where $\beta = 1$, $M = \varepsilon E^2/\rho\langle\sigma\rangle$ and $N = \Delta\sigma/L$, $k_k = (Ra_e/L^3)^{1/3}$, where $Ra_e = \frac{\varepsilon E^2 L^2 \Delta\sigma}{4\nu D_s \rho\langle\sigma\rangle}$ is an electric Rayleigh number (see Baygents & Baldessari (1998)) and $L$ is a characteristic large scale. Interestingly, from Eq. (3.9), we can see a singularity emerges at $\beta = 4$. When $\beta$ is over 4, the exponent becomes negative and $k_{us}$ becomes increases with $\nu$ and $D_s$. This is unreasonable in a physical system under a certain equilibrium state, since a smaller $\nu$ should provide a larger $k_{us}$. In other words, if in a forced turbulent system with $\beta > 4$, the system could be unstable and crash, till a new equilibrium state with $\beta < 4$ is reestablished. Thus, in an equilibrium system, we have to claim that $\beta \leq 4$, which is again coincident with Zhao and Wang (2021). When $\beta = 4$, $k_{us}$ is infinite and unachievable.

In the investigation of Zhao and Wang (2021), they also advanced a small length scale $l_K$. It should be noted, $k_{us}$ is not the wavenumber corresponding to $l_K$. $l_K$, or the corresponding wavenumber $k_K$ can be obtained by merely $F_s(k_K) = D_u(k_K)$, while $k_{us}$ is obtained from Eq. (3.8). In EK turbulence where $\beta = 1$, the former has $k_K \sim Ra_e^{5/12}$ and the latter has $k_{us} \sim Ra_e^{1/3}$.

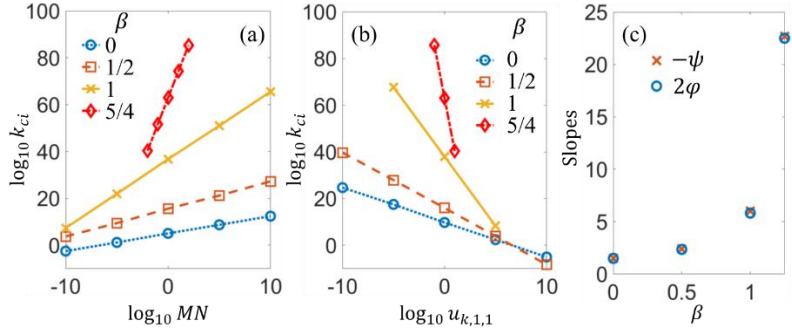

Figure 5 Critical wavenumber $k_{ci}$ varies with $MN$ and $u_{k,1,1}$ for different $\beta$. (a) $k_{ci}$ vs $MN$ at $u_{k,1,1} = 1$. (b) $k_{ci}$ vs $u_{k,1,1}$ at $MN = 1$. (c) Comparison between $2\varphi$ and $-\psi$.

### 3.3 Influence of the ith solution

Eq. (3.1) is a third-order equation of $u_{k,i}$. It has three solutions in total, including one real solution and two complex solutions. In the investigation of Alam et al (2019), they solved the fifth-order equation (Eq. (4.2) in their paper) but only kept the real solution. However, as explained above, when we think about $E_u = k^{-1}(u_k \cdot u_k^*)$, the complex solutions of $u_k$ can also lead to real $E_u$. Thus, it is necessary to revisit the roles of complex solutions of $u_k$ in the turbulent transport. It should be noted that Eq. (4.2) in Alam et al (2019) has 5 solutions with only three of them being independent. The independent solutions in their investigation are exactly the same as our numerical calculation with Eq. (3.1) with $\beta = 0$.

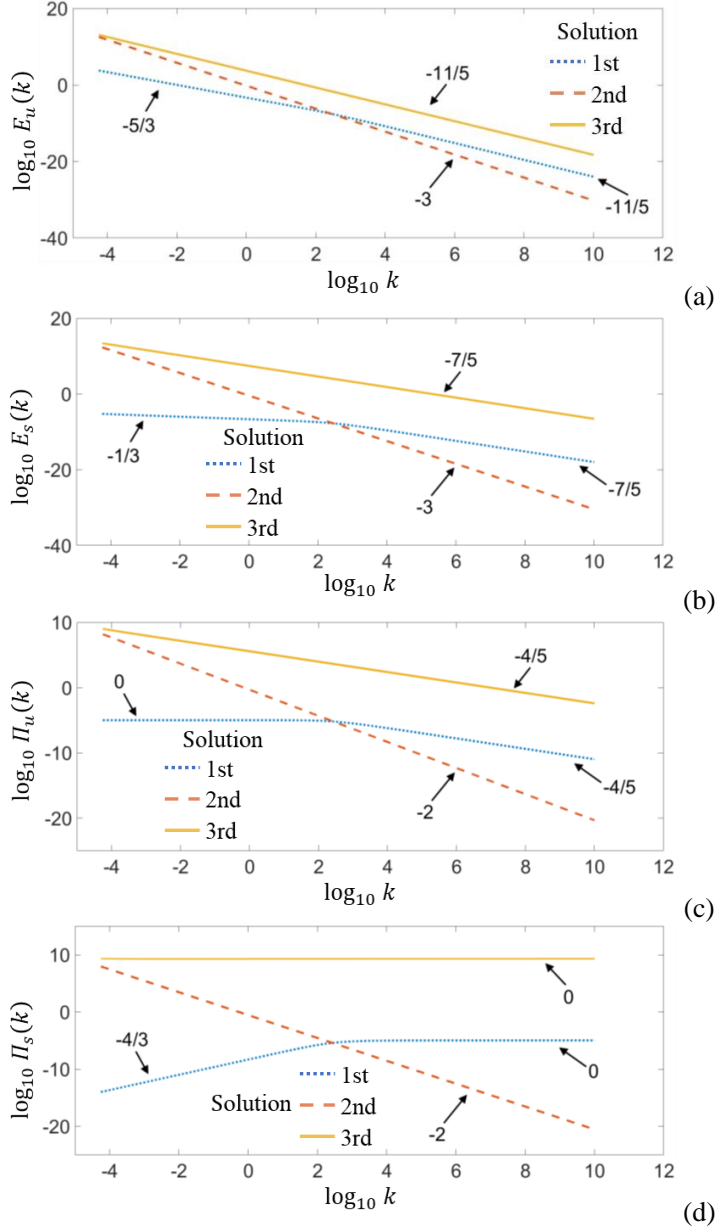

Figure 6 Influence of the ith solution, where $M = N = 1$, $\beta = 0$. (a) Power spectra of kinetic energy $E_u$. (b) Power spectra of scalar variance $E_s$. (c) Flux of kinetic energy $\Pi_u$. (d) Flux of scalar variance $\Pi_s$.

From Figure 6, it can be seen the 2nd and 3rd solutions both exhibit a single slope in the VFD subrange. For the 2nd solution, when $\beta < 3/2$, there is no constant flux of kinetic energy and scalar variance. As an example, at $\beta = 0$, $\xi_u = \xi_s = -3$ (Figure 6(a, b)) and $\lambda_u = \lambda_s = -2$ (Figure 6(c, d)). However, when $\beta \geq 3/2$, the 2nd solution exhibit a single CEF subrange, where $\xi_u = -5/3$ and $\lambda_u = 0$. $\xi_s$ and $\lambda_s$ can be calculated by Eq. (3.3) and (3.4). While for the 3rd solution, when $\beta < 3/2$, there is always a CSF subrange with $\lambda_s \approx$ 0, where $\xi_u$, $\lambda_u$ and $\xi_s$ can be calculated from Eq. (3.5)-(3.7). For instance, when $\beta = 0$, $\xi_u = -11/5$, $\xi_s = -7/5$ and $\lambda_u = -4/5$ respectively. However, when $\beta \geq 3/2$, $\lambda_s$ gradually departs from zero. The flux of kinetic energy and scalar variance become non-constant in the investigated wavenumber region.

The investigations on the additional solutions show some unpredicted results that may explain some early experimental observations. For instance, in stratified turbulence ($\beta = 0$) in the atmosphere, a -3 slope of the kinetic energy spectrum has been experimentally observed by Nastrom and Gage (Nastrom et al. 1984, Nastrom & Gage 1985) on synoptic scale. A later investigation by numerical simulation (Kitamura & Matsuda 2006) shows both the spectra of kinetic energy and potential energy (equivalent to density variance) have -3 slopes. The researchers focus on how the -3 spectra are generated (e.g. upscale or downscale cascade), but do not explain why there should be -3 spectra intrinsically. This investigation indicates the -3 spectra of kinetic energy and scalar variance are direct consequences of an inherent solution of the conservation equation (Eq. (2.15) or (3.1)). It must appear if some transitional conditions are satisfied.

## 4. Asymptotic analysis

Eq. (2.15) can be equivalently rewritten as

$$\frac{du_k}{dk} = -\frac{u_k}{k}\frac{1+(3-2\beta)A(k)}{3+5\,A(k)} \tag{4.1}$$

where $A(k) = k^{2-\beta}\frac{u_k^2}{MN}$. Let $u_k \sim k^{f(\beta)}$, we have

$$f(\beta) = \frac{d\ln u_k}{d\ln k} = -\frac{1+(3-2\beta)A(k)}{3+5\,A(k)} \tag{4.2}$$

On one hand, when $A(k)$ approaches 0, $f(\beta) = -1/3$, which in turn leads to $\xi_u = -5/3$ according to Eq. (2.13a). This is exactly what we found in the CEF subrange. On the other hand, when $A(k)$ approaches infinity, we have $f(\beta) = (2\beta - 3)/5$, which in turn leads to $\xi_u = (4\beta - 11)/5$. This is consistent with Eq. (3.4) in the CSF subrange.

When $\beta = 2$, $f(\beta) = 0$ as can be seen from the numerical simulation above. Accordingly, we have $A(k) = (MN)^{-1}$ and $f(2) = -\frac{1-(MN)^{-1}}{3+5\,(MN)^{-1}}$. It is interesting to see, if $MN = 1$, $f(2)$ is exactly zero. This explains why in the numerical simulation, we can have special and singular scaling exponents at $\beta = 2$.

To get the singularity, two conditions must be satisfied simultaneously. One is $f(\beta) = (\beta - 2)/2$, the other is $MN = (16 - 9\beta)/(3\beta - 8)$. For each $\beta$, we can find a $MN$ to reach the singular scaling exponents which are $\xi_u = \beta - 3$, $\xi_s = -3$, $\lambda_u = \frac{3}{2}\beta - 2$ and $\lambda_s = \frac{1}{2}\beta - 2$. It should be noted that in practical applications, it is almost impossible to reach the singular scaling exponents, since $MN$ cannot be exactly $(16 - 9\beta)/(3\beta - 8)$. Accordingly, the subrange returns to CSF at the $\beta$.

## 5. Discussions

To this end, four different cascade processes have been predicted theoretically in the scalar-based forced turbulence, with either constant or non-constant fluxes of kinetic energy and scalar variance. The results are summarized in Table 1.

Table 1. Cascade processes in the scalar-based forced turbulence

| Subrange | Features | |
|---|---|---|
| | $\Pi_u$ | $\Pi_s$ |
| Inertial | Constant | Constant |
| CEF | Quasi-constant | Non-constant |
| CSF | Non-constant | Quasi-constant |
| Dissipation | Non-constant | Non-constant |
| VFD (2nd solution) | Non-constant | Non-constant |

In this investigation, defining the property of $\beta$ is difficult. From the spectral format of Eq. (2.4c), it is appropriate to define $\beta$ as any real number no more than 4, including the fractional number and even negative number if physically appropriate. It is unable to answer this question here and we hope to leave it for future research. Applying real $\beta$ provides an effective tool to classify the turbulence driven by a scalar-based volume force, since the same or similar $\beta$ in different turbulent systems reserve the same or similar transport process of kinetic energy and scalar variance. For instance, if let $\beta = 1$, $\xi_s = -7/3$ in the CEF subrange which is exactly coincident with the density spectrum in magnetohydrodynamic (MHD) turbulence predicted by Batchelor in terms of pressure fluctuations (Batchelor 1951, Montgomery et al. 1987, Biskamp 2003). Interestingly, if we slightly change $\beta$ to 0.9, we have $\xi_s = -2.13$ in the CEF subrange and $\xi_s = -1.76$ in the CSF subrange. These values are tightly close to the findings (-2.2 and -1.7) of Kowal et al (2007) on the density fluctuations in MHD turbulence. Therefore, $\beta \approx 1$ can characterize both EK and MHD turbulences. Another example is when $\beta = -1$, $\xi_u = -3$ and $\xi_s = -1$ in the CSF subrange are approaching exponents in the elastic turbulence (Groisman & Steinberg 2000), which is intrinsically a chaotic flow (Steinberg 2019). Instead of seeking the solution to each type of turbulence, researchers can numerically solve the conceptual model in Eqs. (2.1a-c) with $\beta$ for general solutions.

## 6. Conclusions

In this investigation, we numerically studied the transport of kinetic energy and scalar variance in turbulence driven by scalar-based volume force. In this type of turbulence, the scalar field is strongly coupled with the velocity field and dominated by active transport. A conservative equation relying on the fluxes of kinetic energy and scalar variance has been established. The equation has three solutions, including one real solution and two complex solutions. Based on the solutions, a comprehensive cascade picture has been established for this type of turbulence.

The investigations on the real solution indicate the turbulence has 4 types of cascade processes, depending on whether the fluxes of kinetic energy and scalar variance are constant or not. In the inertial subrange, both the fluxes of kinetic energy and scalar variance are constant. In the CEF and CSF subranges, only the flux of kinetic energy or scalar variance is quasi-constant. And in the dissipation subrange, both the fluxes of kinetic energy and scalar variance are non-constant.

The scaling exponents in the power spectra of kinetic energy and scalar variance strictly depend on the order of derivatives, i.e. $\beta$. In the CEF subrange, $\xi_u$ is always -5/3, while $\xi_s = -(6\beta + 1)/3$. In the CSF subrange, $\xi_u = (4\beta - 11)/5$ and $\xi_s = -(2\beta + 7)/5$ which are both consistent with the theory of Zhao and Wang (2021). Relying on $\beta$, the transport of kinetic energy and scalar variance can be distinguished as four cases. (1) When $\beta < 3/2$ (except $\beta = 2/3$), the CEF and CSF subranges both exist in the investigated wavenumber region, and the former locates on the lower wavenumber side of the latter. If $\beta = 2/3$, the CEF and CSF subrange overlap and form a new inertial subrange, where both $\xi_u$ and $\xi_s$ equal to -5/3. (2) When $3/2 \leq \beta < 2$, only the CEF subrange is observed. (3) At $\beta = 2$, if $MN = 1$, special and singular exponents of $\xi_u = -1$, $\xi_s = -3$, $\lambda_u = 1$, and $\lambda_s = -1$ can be found. Otherwise, if $MN \neq 1$, the CSF subrange is observed. (4) When $2 < \beta \leq 4$, only the CSF subrange is observed. Note, $\beta = 4$ is a critical point for the equilibrium state to be established. Beyond this, the equilibrium state of the turbulence system could be crashed.

We also examined the influence of $MN$ and the initial value of the real solution $u_{k,1,1}$ on the cross over wavenumber between the CEF and CSF subranges, when $\beta < 3/2$ (except $\beta = 2/3$). The results indicate $k_{ci} \sim (MN)^{\varphi} u_{k,1,1}{}^{\psi}$, with $\varphi > 0$ and $\psi < 0$. It is interesting that $\psi \approx -2\varphi$. The larger the $\beta$ is, the larger $|\psi|$ and $\varphi$. Another microscale $k_{us} = (MN/4\nu D_s)^{\frac{1}{4-\beta}}$ is also predicted if the CSF subrange is present. On $k_{us}$, the energy feeding rate and scalar feeding rate are simultaneously balanced by the energy and scalar dissipation respectively. The formula of $k_{us}$ prevents $\beta$ to be larger than 4.

The results above are all for the first solution. For the 2nd solution, when $\beta < 3/2$, there is no constant flux of kinetic energy and scalar variance. When $\beta \geq 3/2$, a single CEF subrange is predicted. While for the 3rd solution, when $\beta < 3/2$, there is always a CSF subrange with $\lambda_s \approx 0$. However, when $\beta \geq 3/2$, $\lambda_s$ gradually departs from zero, and the flux of kinetic energy and scalar variance become non-constant.

This paper aims to establish a comprehensive picture of the cascade process in the forced turbulence driven by scalar-based volume force and the accompanied active transport process, within a unified theoretical frame. In the meanwhile, we hope this investigation can significantly deepen our understanding of the transport of kinetic energy and scalar variance, e.g. in stratified turbulence, electrokinetic turbulence and other physical systems like condensed matter physics. The scalar is not restricted to any known physical quantities, e.g. temperature, density, electric conductivity and permittivity etc, but also includes broad scalar quantities like information.

**Acknowledgement** The investigation is supported by the National Nature Science Foundation No. 11672229.